\documentclass[aps,prc,floatfix,twocolumn,nofootinbib,superscriptaddress]{revtex4-1}

\usepackage[]{graphicx}
\usepackage{color}
\usepackage{subfigure}
\usepackage{amsmath,amssymb,amsfonts}

\usepackage{slashed}

\begin{document}

\title{Exotic hybrid pseudopotentials at finite temperature and chemical potential}

\author{Le Zhang}
\affiliation{College of Physics and Electronic Science, Hubei Normal University, Huangshi, 435002, China}

\author{Fei-Yang Cai}
\affiliation{College of Physics and Electronic Science, Hubei Normal University, Huangshi, 435002, China}

\author{Xun Chen}
\email{chenxunhep@qq.com}
\affiliation{School of Nuclear Science and Technology, University of South China, Hengyang 421001, China}

\date{\today}

\begin{abstract}
Using gauge/gravity duality, we study the exotic hybrid pseudopotentials at finite temperature and chemical potential. The $\Sigma$ hybrid meson can be described by a model including an object called ``defect'' on a string linking the quark and antiquark. It was first proposed by Andreev and perfectly described the $\Sigma_u^-$ hybrid potential at zero temperature and chemical potential. In this paper, we would like to extend this model to finite chemical potential and compare the separate distance and pseudopotentials of $\Sigma_g^+ $ and $\Sigma_u^-$. Unlike $\Sigma_g^+$ ground state, the $\Sigma_u^-$ hybrid pseudopotentials no longer behave as Coulomb-like at short distances. In addition, temperature and chemical potential have a significant impact on the $\Sigma_u^-$ hybrid pseudopotentials. The screen distances and hybrid pseudopotentials of $\Sigma_u^-$ significantly decrease with the increase of temperature and chemical potential. At last, we draw the melting diagram of $\Sigma_g^+ $ and $\Sigma_u^-$ in the $T- \mu$ plane, and confirm that the quark-antiquark pair in $\Sigma_u^-$ excited state is easier to melt than in $\Sigma_g^+$ ground state.
\end{abstract}

\maketitle
%
\begin{widetext}

\section{Introduction}
The quark-antiquark potential is one of the best-studied quantities in QCD due to its importance in spectroscopy and phenomenology. In the early stages of establishing QCD, it was related to the large time behavior of Wilson loop expectation value~\cite{Fischler:1977yf, Brown:1979ya}. Wilson loop expectation values can be computed in lattice QCD. Thus lattice QCD has been regarded as the main tool for studying quark-antiquark potential, and their resulting potential agreement with phenomenology~\cite{Bali:1997am, Baker:1997bg, Bali:2000gf, Takahashi:2002bw, Ratti:2005jh, Bicudo:2007xp, Luscher:2010iy, Juge:2002br, Koma:2017hcm, Muller:2019joq, Brambilla:2021wqs, Schlosser:2021hed, Schlosser:2021wnr, Schlosser:2022twt, Philipsen:2013ysa,Bala:2020tdt}.
On the other hand, $\rm{AdS/CFT}$ correspondence~\cite{Maldacena:1997re, Gubser:1998bc, Witten:1998qj} is that it opened a new window for dealing with strongly coupling gauge theories. Within the AdS/CFT
correspondence, the expectation values of an operator similar to the Wilson loop can be calculated in~\cite{Maldacena:1998im, Aharony:1999ti}.
As a result, the study of the quark-antiquark potential by using gauge/string duality has received attention.
Ref.~\cite{Maldacena:1998im} is the earliest literature to study the static quark-antiquark potential by using gauge/string duality and discovers that it exhibits Coulombian behavior at short distances. The pseudopotential of quark-antiquark pair was extended in Refs.~\cite{Rey:1998bq, Brandhuber:1998bs, Andreev:2006eh}, and the influence of chemical potential on pseudopotential is discussed in~\cite{Fadafan:2012qy, Lee:2014gna, Chen:2017lsf, Jiang:2022zbt}. It was found that the quark-antiquark pseudopotential at any temperature and chemical potential is Coulomb-like behavior at short distances and linear at long ranges. After that, some extreme conditions (e.g., rotating meson, rotating matter, anisotropy and magnetic field)~\cite{Chakraborty:2012dt, Finazzo:2014rca, Gursoy:2020kjd, Chen:2017lsf, Zhou:2020ssi, Zhou:2021sdy, Zhou:2023qtr}, and some holographic models of multiquark (pseudo)potential~\cite{Andreev:2008tv, Andreev:2015riv, Andreev:2015iaa, Andreev:2019cbc, Andreev:2020xor, Andreev:2021bfg, Andreev:2021eyj, Andreev:2022cax, Chen:2021bkc, Jiang:2022zbt, Yu:2023hzl, Jiang:2023lmj, Liu:2023hoq} are discussed.

Moreover, QCD predicts the existence of hybrid mesons and baryons in which the gluonic field between a quark and antiquark is in an excited state. Thus hybrids containing the excited gluonic field are an important source of information related to QCD. The properties of Hybrid mesons and baryons are predicted by lattice simulations in Refs.~\cite{Isgur:1984bm, McNeile:2002en}. Then the hybrid static potentials have been the object of many calculations in lattice QCD~\cite{Juge:1997nc, Michael:1998tr, Bali:2000gf, McNeile:2002en, Michael:2003xg, Juge:2002br, Bicudo:2008na, Wolf:2014tta, Meyer:2015eta, Reisinger:2017btr, Capitani:2018rox, Schlosser:2021wnr, Schlosser:2021hed}. To our knowledge, there are currently no studies of the pseudopotential of hybrid mesons and baryons in lattice QCD simulations.

As a prelude to discussing the hybrid pseudopotentials, let us briefly consider the quark-antiquark pseudopotential. The pseudopotential,  which is confining for all temperatures, can be extracted from the expectation value of a Wilson loop~\cite{Borgs:1985qh, Manousakis:1986jh}. We can do so by adopting an important example that is a rectangular Wilson loop of size $L \times Y$, oriented in the $x y$-plane. In this case, the expectation value of the Wilson loop is~\cite{Andreev:2012hw}
\begin{eqnarray}
\langle W (L \times Y)\rangle \sim \sum_{n=0}^{\infty} c_n (L) e^{-E_n(L) Y}.
\end{eqnarray}
If the corresponding contribution dominates the sum as $Y \rightarrow \infty$, $E$  represents the pseudopotential and the other $E_n$'s are called hybrid pseudopotentials.

Inspired by the AdS/CFT correspondence, Andreev et al. proposed a model within a five(ten)-dimensional effective string theory, which can originally be used to compute the quark-antiquark (pseudo)potential~\cite{Andreev:2006ct, Andreev:2006eh} and the three-quark (pseudo)potential~\cite{Andreev:2008tv} and exotic states \cite{Andreev:2024orz}. By inserting a new object called ``defect'' on a string linking quark and antiquark, this model has been developed to calculate exotic hybrid potential~\cite{Andreev:2012mc} and pseudopotential~\cite{Andreev:2012hw} of quark-antiquark pair. For an example of $\Sigma_u^-$ (It represents the first excited state of $\Sigma$ meson, and $\Sigma_g^+$ means the ground state of $\Sigma$ meson.), the resulting hybrid potential at zero temperature of Ref.~\cite{Andreev:2012mc} perfectly matches lattice simulations in Ref.~\cite{Co:2021lkc}. In Ref.~\cite{Andreev:2012mc}, it can be observed that the $\Sigma_u^-$ hybrid pseudopotential no longer exhibits Coulomb-like behavior at short distances.

Large Hadron Collider and relativistic heavy-ion collision experiments are able to simulate high-temperature and high-density environments in the laboratory. Research on exotic hybrid pseudopotentials at finite temperature and chemical potential provides theoretical background for these experiments, guiding the observation and interpretation of results. Neutron stars and other dense astrophysical objects may have extreme temperature and chemical potential conditions internally. Studying pseudopotentials under these conditions can provide insights into the state of matter and its interactions within these celestial bodies.

In this paper, we would like to extend hybrid pseudopotential to finite temperature and chemical potential. That is to say, we want to know how the temperature and chemical potential affect the hybrid pseudopotentials. The organization of the paper is as follows: In Sect.~II, a string configuration of exotic hybrid mesons will be briefly introduced. In Sect.~III, we will describe a concrete example of the ten-dimensional effective string theory at finite temperature and chemical potential. This theory was proposed by Andreev and could effectively describe the static potential of $\Sigma_g^+$ and $\Sigma_u^-$ at zero temperature and chemical potential~\cite{Andreev:2012mc, Andreev:2012hw}. We will extend it to finite chemical potential and calculate the hybrid pseudopotential of $\Sigma_u^-$ at finite temperature and chemical potential. Then, we present the numerical results and compare the $\Sigma_u^-$ hybrid pseudopotential with
$\Sigma_g^+$ pseudopotential in Sect.~IV. The last section contains our conclusion and discussion.

\section{ A string configuration of exotic hybrid quark-antiquark pair }
In this section, we first briefly review a string configuration of exotic hybrid mesons.
Common wisdom has it that the ground state energy of a $U$-shape string linking two fermionic sources corresponds to the potential of quark-antiquark pair and the excited strings (fluxes) induce hybrid potentials~\cite{Isgur:1984bm}. The string excitations may be modeled by inserting local objects on the linking string, if we take into account that some excitations remain very narrow along a string. these objects, being called defects,
are just the macroscopic description of some gluonic degree of freedom in strong coupling~\cite{Andreev:2012mc, Andreev:2012hw}.
It should be noted that when defects are inserted, the string embeddings of the world sheets in spacetime become non-differentiable at the locations of these defects.
This model, considering the case of a single defect, was used in Ref.~\cite{Andreev:2012mc} to perfectly reproduce the lattice data of Ref.~\cite{Co:2021lkc} for $\Sigma_u^-$ hybrid potential at zero temperature and was used in Ref.~\cite{Andreev:2012hw} to predict $\Sigma_u^-$ hybrid pseudopotential at finite temperature. In this paper, we will extend it to finite temperature and chemical potential and pay attention to the influence of temperature and chemical potential on exotic hybrid pseudopotentials corresponding to the excited state.

In Fig.~\ref{shiyitu}, we show a possible configuration of a hybrid quark-antiquark pair. In this string configuration, the quark (or anti-quark) located at the boundary points of a five (ten) dimensional space is the endpoint of a fundamental string. These two fundamental strings connect at a defect in internal.
It should be noted that we have set two symmetries to simplify the string configuration based on the delicate analysis of Andreev in Refs.~\cite{Andreev:2012mc, Andreev:2012hw}.
In doing so, this string configuration sits entirely on the $xor$ plane (${y_D}=0$), and the defect $D$ must be on the $r$-axis (${x_D}=0$).
\begin{figure}
\centering
\includegraphics[width=6cm]{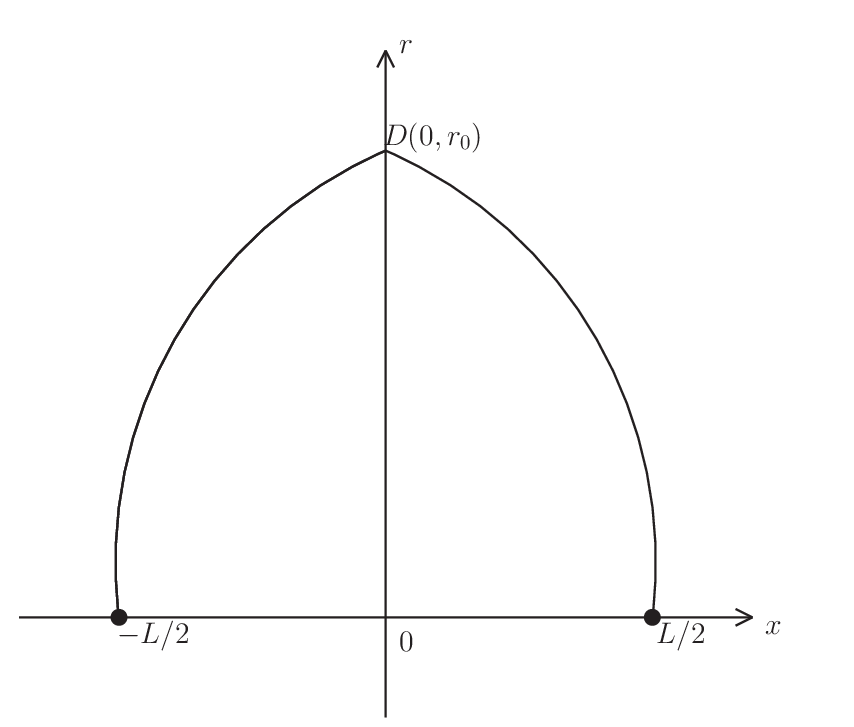}
\caption{A string configuration of the hybrid quark-antiquark pair in the case of a single symmetrical defect. The quark and antiquark are on the $x$-axis at $x=L/2$ and $x=-L/2$, respectively. The defect $D$ is located on the $xor$ plane. If there is no defect $D$, the string configuration can be described as the quark-antiquark pair corresponding to a ground
state connected by a U-shape string.}
\label{shiyitu}
\end{figure}

In such a string configuration with a single symmetrical defect(seeing Fig.~\ref{shiyitu}), the action not only includes two standard Nambu-Goto actions but also the contribution arising from the defect. It is thus,
\begin{eqnarray}
\label{stringaction}
S=\sum_{i=1}^2 S^{\rm{(N G)}}_i +S^{\rm{(def)}}
\end{eqnarray}
where $S^{\rm{(N G)}}_i$ denotes the Nambu-Goto action of the string linking the $i$ fermionic source(quark or antiquark) and the defect D.
And $S_{\rm{def}}$ is seen as the action for the defect, which may be expressed as~\cite{Andreev:2012mc, Andreev:2012hw},
\begin{eqnarray}
\label{stringaction1}
S^{\rm{(def)}}(r_0) = {T}\; \cdot \; \mathcal{E}^{\rm{(def)}}(r_0),
\end{eqnarray}
here $\mathcal{E}^{\rm{(def)}}(r_0)$ is taken to be the effective potential for the defect $D$. It might be noted that its specific form requires the framework of the AdS/QCD duality.

\section{Hybrid pseudopotentials at finite temperature and chemical potential }
Now we will use $\Sigma_u^-$ hybrid meson as a concrete example to compute the hybrid pseudopotentials of quark-antiquark pair based on the string configuration mentioned in the previous section. As presented in Refs.~\cite{Andreev:2012mc, Andreev:2012hw}, the single defect lives in a ten-dimensional space. In ten-dimensional space, there is nothing but a brane-antibrane pair~\cite{Witten:1998xy}. At leading order in $\alpha'$, the brane action is seems to assume that $S^{\rm{(def)}} \sim \tau_5 \int d^6 x \sqrt{g^{(6)}}$ because that the branes discussed in this model are fivebranes~\cite{Witten:1998xy}. Here $\tau_5$ is a brane tension and $x^{i}$ are the world-volume coordinates. Then the ten-dimensional geometry at finite temperature and chemical potential for subsequent applications possibility is taken as~\cite{Andreev:2007zv, Andreev:2007rx, Andreev:2015riv},
\begin{eqnarray}
d s^{2}=\mathcal{R}^{2}\varpi(r)\left(f(r)d t^{2}+d \vec{x}^{2}+ f^{-1}(r) d r^{2}\right)+\mathrm{e}^{-\mathfrak{s} r^{2}} g_{a b}^{(5)} d \omega^{a} d \omega^{b},
\\
\varpi(r)=\frac{e^{\mathfrak{s} r^{2}}}{r^{2}},\;\;\;\;f(r)=1-(1 + Q^2) \left(\frac{r}{r_h}\right)^{4}+ Q^{2} \left(\frac{r}{r_h}\right)^{6}.\nonumber
\end{eqnarray}
This is a one-parameter deformation of Euclidean $AdS_5$ space with constant radius $\mathcal{R}$ and a 5-dimensional compact space(sphere) $\textbf{X}$ whose coordinates are $\omega^a$. Here $\mathfrak{s}$ is a one-parameter deformation that can be fixed from quark-antiquark potential corresponding to the ground state at vanishing temperature and chemical potential, and $Q$ is the black hole charge. The parameter $r_h$, representing the position of the black hole horizon, is related to the Hawking temperature of the black hole. That identified with the temperature of a dual gauge theory can be defined as,
\begin{eqnarray}
T=\frac{1}{4 \pi}\left|\frac{d f}{d r}\right|_{r=r_{h}}=\frac{1}{\pi r_{h}}\left(1-\frac{1}{2} Q^{2}\right),
\end{eqnarray}
here $Q$ is in the range $0 \leq Q \leq \sqrt{2}$ to ensure positive temperature. The relationship between the chemical potential $\mu$ and $Q$ is $\mu={Q}/{r_{h}}$ in this paper. Note that the chemical potential here is not the real quark (or baryon) chemical potential in QCD but corresponds to the R-charge of $N = 4\; \rm{SYM} $ theory.

For subsequent computations, we assume that the fivebranes wrap on $\mathbf{R} \times \textbf{X}$, with $\mathbf{R}$ along
the y-axis in the deformed ${AdS}_{5}$, the action $S^{\rm{(def)}}$ contributed by an extra item arising from the defect $D$ in Eq.~(\ref{stringaction1}) may be represented as,
\begin{eqnarray}
S^{\rm{(def)}} = \mathfrak{m} \mathcal{R} \int d t \frac{\mathrm{e}^{- 2 \mathfrak{s} r^{2}}}{r} \sqrt{f(r)},
\end{eqnarray}
with $\mathfrak{m} \sim \tau_5 \int d^5 \omega \sqrt{g^{(6)}}$. The extra term of the effective pseudopotential comes from the contribution of the defect $D$ and may be written as the following form,
\begin{eqnarray}
\mathcal{E}^{\rm{(def)}}(r_0) = \frac{S^{\rm{(def)}}(r_0)}{T} = \mathfrak{m} \mathcal{R} \frac{\mathrm{e}^{- 2 \mathfrak{s} r_0^{2}}}{r_0} \sqrt{f(r_0)}.
\end{eqnarray}

Next, we need to return to the calculation of the standard Nambu-Goto action $S^{\rm{(N G)}}_i$ of the string linking the $i$- fermionic source and the defect $D$.
If we choose the static gauge $\xi^{1}=t$ and $\xi^{2}=r$, then the boundary conditions for $x(r)$ become,
\begin{eqnarray}
x^{(1)}(0)=-\frac{L}{2}, \;\;\;\; x^{(2)}(0)=\frac{L}{2}, \;\;\;\; x^{(i)}(r)=0
\end{eqnarray}
By using the static gauge $t=\tau, \sigma=x$, the Nambu-Goto action $S^{\rm{(N G)}}_i$ of one of the two $i$-strings linking the two fermionic source(quark or antiquark) and the defect $D$ is
\begin{eqnarray}\label{sng}
S^{\rm{(N G)}}_i=\frac{1}{2 \pi \alpha^{\prime}} \int d \tau d \sigma \sqrt{\operatorname{det} g_{\alpha \beta}},
\end{eqnarray}
with
\begin{eqnarray}
g_{\alpha \beta}=G_{\mu \nu} \frac{\partial x^{\mu}}{\partial \sigma^{\alpha}} \frac{\partial x^{\nu}}{\partial \sigma^{\beta}},
\end{eqnarray}
where $x^{\mu}$ and $G_{\mu \nu}$ are the target space coordinates and the metric respectively, and $\sigma^{\alpha}$ with $\alpha = 0, 1$ parameterize the world sheet.

The Nambu-Goto action of the left string in Fig.~\ref{shiyitu} (which we refer to as $S^{\rm{(N G)}}_1$) can be written as,
\begin{eqnarray}
S^{\rm{(N G)}}_{1}=\mathfrak{g}\,{T} \int_{-\frac{L}{2}}^{0} d x \varpi(r) \sqrt{f(r)+\left(\partial_{x} r\right)^{2}},
\end{eqnarray}
where $\mathfrak{g}=\frac{R^2}{2 \pi \alpha^{\prime}}$ is defined for convenience. Recalling that this string configuration is symmetrical~\cite{Andreev:2012mc, Andreev:2012hw}, we can easily get this relationship $S^{\rm{(N G)}}_2 = S^{\rm{(N G)}}_1$ ($S^{\rm{(N G)}}_2$ means the Nambu-Goto action of the right string in Fig.~\ref{shiyitu}).

Now the Lagrangian density $\mathcal{L}_1$ of the left string is identified as
\begin{eqnarray}
\mathcal{L}_1=\varpi(r) \sqrt{f(r)+\left(\partial_{x} r\right)^{2}}.
\end{eqnarray}
Therefore we get the first integral of Euler-Lagrange equations,
\begin{eqnarray}
\mathcal{H}_1(r)&=&
\mathcal{L}_1 - \frac{\partial \mathcal{L}_1 }{\partial (\partial_x r )}
\nonumber\\&=&
\frac{\varpi(r) f(r)}{\sqrt{f(r)+\left(\partial_{x} r\right)^{2}}},
\end{eqnarray}
which is a constant.

At the endpoint $r_0$ where the defect $D$ is located, we have $\left.\partial_{x} r\right|_{r=r_{0}}=\tan \alpha$ and the first integral of the left string at $r_0$ can be written as,
\begin{eqnarray}
\mathcal{H}_1(r_0)=\frac{\varpi(r_0) f(r_0) }{\sqrt{f(r_0)+\tan^{2} \alpha}}.
\end{eqnarray}
Then $\partial_{r} x$ can be solved as,
\begin{eqnarray}
\partial_{r} x = \sqrt{\frac{\varpi(r_0)^2 f\left(r_{0}\right)^{2}}{\left(f\left(r_{0}\right)+\tan ^{2} \alpha\right) \varpi(r)^2 f(r)^{2}-f(r) \varpi(r_0)^2 f(r_0)^{2}}}.
\end{eqnarray}
It should be noted here that the left string at the endpoint $r_0$ is not continuous due to the presence of the defect $D$, that is, $\tan \alpha \neq 0$ with $\alpha >0$. So is the right string due to the symmetry of the string configuration.

Based on the above, the separate lengths of the quark and antiquark can be expressed as,
\begin{eqnarray}
\label{separate length1}
L=2 \int_{0}^{r_{0}} \partial_{r} x d r
= 2 \int_{0}^{r_0} \,d r\,\sqrt{\frac{\varpi(r_0)^2 f(r_0)^2}{(f(r_0)+\tan ^2 \alpha) \varpi(r)^2 f(r)^2-f(r) \varpi(r_0)^2 f(r_0)^2}}.
\end{eqnarray}

As for $E=\frac{S}{T}$, the potential contributed by the strings is,
\begin{eqnarray}
\label{PECBS}
E^{\rm{(N G)}} = 2\mathfrak{g}\int_{0}^{r_{0}} d r \varpi(r)
\sqrt{1+f(r)(\partial_{r} x)^{2}}.
\end{eqnarray}
Subtracting the divergent term $\mathfrak{g} \int_0^{\infty} dr \frac{1}{r^2}$, we can get the potential $E^{\rm{(N G)}}$ contributed by the two fundamental strings,
\begin{eqnarray}
\label{potentials0}
E^{\rm{(N G)}} = \mathfrak{g}\left[\int_{0}^{r_{0}} d r \left(\varpi(r)
\sqrt{1+f(r)(\partial_{r} x)^{2}}-\frac{1}{r^2}\right)-\frac{1}{r_0}\right]+C,
\end{eqnarray}
where $C$ is a normalization constant. It is noted that the above formulas, Eq.~(\ref{separate length1}) and Eq.~(\ref{potentials0}), can be reduced to the separate length and potential of quark-antiquark pairs in the ground state when $\left.\partial_{x} r\right|_{r=r_{0}}=\tan \alpha = 0$~\cite{ Andreev:2006ct, Andreev:2007rx}. Of course, they are used to calculate the relevant results of a quark-antiquark pair in the ground state for comparison with those in the excited state.

Recalling the string configuration for exotic hybrid quark-antiquark pair, the regularized pseudopotentials at finite temperature and chemical potential should be contributed by the two $i$-strings and the defect $D$. So the exotic hybrid pseudopotential corresponding to the hybrid quark-antiquark pair may be written as,
\begin{eqnarray}
\label{potentials1}
E &=& \mathcal{E}^{\rm{(def)}}(r_0) + E^{\rm{(N G)}}
\nonumber\\
&=& 2\mathfrak{g}\left[\int_{0}^{r_{0}} d r \left(\varpi(r)
\sqrt{1+f(r)(\partial_{r} x)^2}-\frac{1}{r^2}\right)-\frac{1}{r_0} + \mathbf{\kappa} \frac{\mathrm{e}^{- {2} \mathfrak{s} r_0^{2}}}{r_0} \sqrt{f(r_0)}\right]+C,
\end{eqnarray}
where $\mathbf{\kappa} = \frac{\mathfrak{m} \mathcal{R}}{2\;\mathfrak{g}}$ is a dimensionless parameter.

Since there is no net force acting on the defect D in equilibrium~\cite{Andreev:2015riv, Andreev:2020xor, Chen:2021bkc, Jiang:2023lmj}, the equation of force balance at the string endpoint $r_0$ leads to,
\begin{eqnarray}
\varpi(r_{0}) \frac{\tan\alpha}{\sqrt{f(r_0)+ \tan^{2}\alpha}}+\left( \kappa \frac{e^{ - {2} \mathfrak{s} r_0^2}}{r_0} \sqrt{f(r_0)}\right)^{\prime}=0.
\end{eqnarray}
So we can easily obtain the following,
\begin{eqnarray}
\label{force balance2}
\tan^{2} \alpha = - f(r_0)+\left[\frac{1}{f(r_0)}
-\kappa^2 \mathrm{e}^{- 6 \mathfrak{s} r_0^{2}}\left(1 +4 \mathfrak{s} r_0^{2} - \frac{r_0 f^{\prime}(r_0)}{2 f(r_0)}\right)^2\right]^{-1}.
\end{eqnarray}
Note that the above equation can be simplified as $\tan^{2} \alpha =0$ if $\kappa$ is set to zero. It means that the two fundamental strings are continuous and differentiable at the endpoints $r_0$. Meanwhile, the exotic hybrid pseudopotential in Eq.~(\ref{potentials1}) is reduced to the result of the ground state. Additionally, if we turn off the chemical potential effect and the temperature effect by setting $Q = 0$ and $T = 0$, these results in Ref.~\cite{Andreev:2012mc} are reproduced. In Fig.~{\ref{figlattice}}, we repeat the energy levels $\Sigma_g^+$ and $\Sigma_u^-$ without temperature and chemical potential in Ref.~\cite{Andreev:2012mc} which fits the lattice data in Ref.~\cite{Co:2021lkc}. The curve corresponding to $\Sigma_g^+$ ground state is calculated by Eq.~(\ref{separate length1}) and Eq.~(\ref{potentials0}). So that we can get $\mathfrak{g}= 0.176$, $\mathfrak{s}= 0.44\, {\rm GeV^2}$, $C = 0.71\, {\rm GeV }$. The remaining free parameter $\mathbf{\kappa}= 2000$ can be determined by fitting the curve of $\Sigma_u^-$ potential calculated by Eq.~(\ref{separate length1}) and Eq.~(\ref{potentials1}) when $ \kappa \neq 0$. It indicates that the potential of $\Sigma_g^+$ ground state and $\Sigma_u^-$ excited state at vanishing temperature and chemical potential can be well described by the string configuration in Fig.~{\ref{shiyitu}}.

\begin{figure}
\centering
\includegraphics[width=7cm]{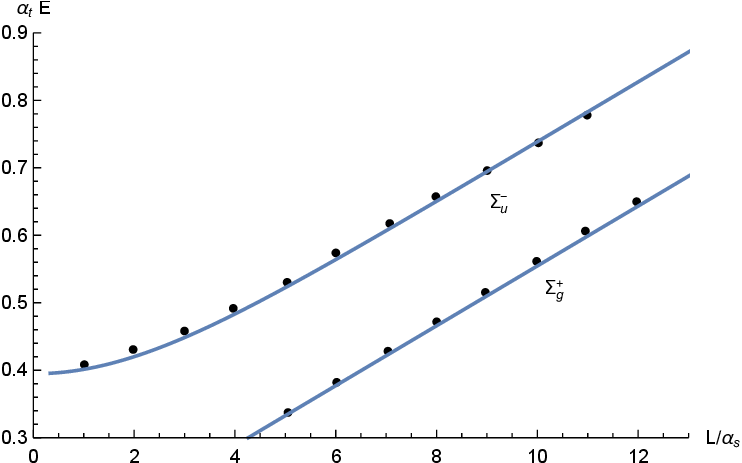}
\caption{Energy levels $\Sigma_g^+$ and $\Sigma_u^-$ without temperature and chemical potential. These curves are our calculation potentials corresponding to $\Sigma_g^+$ ground state and $\Sigma_u^-$ excited state respectively, which reproduce the results from Ref.~\cite{Andreev:2012mc}. The dots represent the lattice data from Ref.~\cite{Co:2021lkc}. Here $\alpha_s = 0.2 \, {\rm {fm}}$ and $\alpha_t \approx 0.41 \, {\rm {fm}} $. }
\label{figlattice}
\end{figure}
At this point in the discussion, we have provided the analytical formula for the pseudopotentials of an exotic hybrid quark-antiquark pair at finite temperature and chemical potential based on the string configuration mentioned in the previous section. In the next section, we will use $\Sigma_u^-$ hybrid meson as an example for discussing the temperature effect and the chemical potential effect on the exotic hybrid pseudopotentials of the quark-antiquark pair.

\section{Numerical results}
In this section, we will present the numerical results based on the analytical calculations in the previous section and will discuss the effects of temperature and chemical potential on the separate distance and the pseudopotentials of an exotic hybrid quark-antiquark pair, respectively. To better analyze the exotic hybrid pseudopotentials, we will compare the examples of $\Sigma_u^-$ and $\Sigma_g^+$.

Firstly, we turn off the chemical potential effect and only consider the contribution of temperature. In Fig.~{\ref{fig2}}, we calculate the separate distance $L$ of $\Sigma_g^+$ and $\Sigma_u^-$, as the function of $r_0$, for different values of temperature $T$. It can be found that the quark-antiquark pair in both cases melts with the increase in temperature. The red curves stand for the melted form, and their maximum value indicates the screening distance of the quark-antiquark pair. Compared to $\Sigma_g^+$ (about $0.132 \; {\rm GeV}$), the melted temperature of $\Sigma_u^-$ drops to about $0.11\; {\rm GeV}$. It means that $\Sigma_u^-$( which is the first $\Sigma$ excited state), as a hybrid meson, is more easily melted than $\Sigma_g^+$ which stands for the string ground state.
\begin{figure}
\centering
\subfigure[.\, Separate distance of $\Sigma_g^+$ ground state]{
\label{fig2a}
\includegraphics[width=0.45\linewidth]{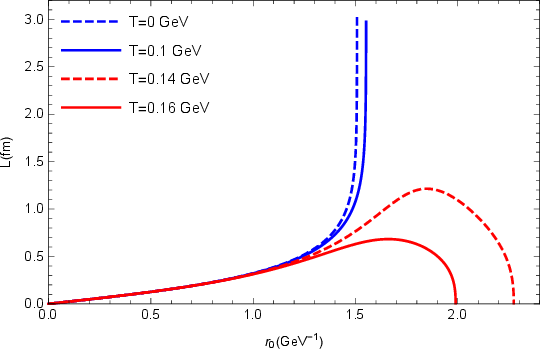}}
\subfigure[.\, Separate distance of $\Sigma_u^-$ excited state]{
\label{fig:subfig:2b}
\includegraphics[width=0.45\linewidth]{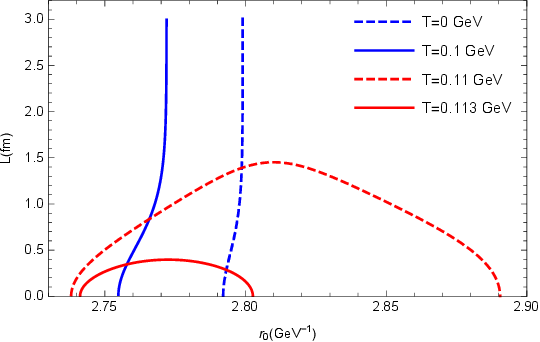}}
\caption{The separate distance $L$ corresponding to $\Sigma_g^+$ and $\Sigma_u^-$ as a function of $r_0$ for different values of $T$ without chemical potential. (a) corresponds to $\Sigma_g^+$ ground state at $\mathfrak{g}= 0.176$, $\mathfrak{s}= 0.44 {\rm GeV^2}$, $C = 0.71 {\rm GeV }$; (b) corresponds to $\Sigma_u^-$ excited state at $\kappa = 2000$. }
\label{fig2}
\end{figure}
The regularized pseudopotential of $\Sigma_g^+$ and $\Sigma_u^-$ as a function of separate distance $L$ for different values of temperature is shown in Fig.~{\ref{fig3}}. At short distances (about $L < 0.6 \; \rm{fm}$), the pseudopotential of $\Sigma_g^+$ at any temperature is Coulomb-like, and the $\Sigma_u^-$ hybrid pseudopotential for any temperature behaves as the quadratic term function of separate distance $L$. At large distances, both $\Sigma_g^+$ pseudopotential and $\Sigma_u^-$ hybrid pseudopotential exhibit a linear behavior. In addition, we found that the temperature has a minor influence on the $\Sigma_g^+$ pseudopotential, especially with almost no effect at small $L$. But the temperature effect on $\Sigma_u^-$ hybrid pseudopotential is significant, whether at large or small separate distance $L$.

\begin{figure}
\centering
\subfigure[.\,$\Sigma_g^+$ pseudopotentials ]{
\label{fig3a}
\includegraphics[width=0.45\linewidth]{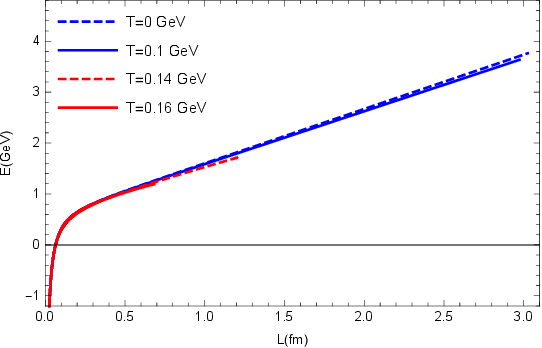}}
\subfigure[.\,$\Sigma_u^-$ hybrid pseudopotentials]{
\label{fig:subfig:3b}
\includegraphics[width=0.45\linewidth]{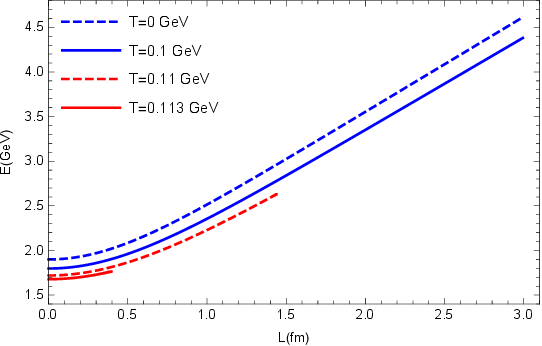}}
\caption{ The regularized pseudopotentials of quark-antiquark
pair corresponding to $\Sigma_g^+$ ground state and $\Sigma_u^-$ excited state as a function
of separate distance $L$ for different values of temperature $T$ at vanishing chemical potential.}
\label{fig3}
\end{figure}
Next, we investigate the behavior of separate distance and pseudopotential at finite temperature and chemical potential in Fig.~{\ref{fig4}} and Fig.~{\ref{fig5}}. Here we take the lower temperature as the fixed value $T = 0.1\, {\rm GeV}$. The separate distance as a function of $r_0$ for different values of chemical potential $\mu$ is shown in Fig.~{\ref{fig4}}. The behaviors of the separate distance of $\Sigma_u^-$ are similar to that of $\Sigma_g^+$ qualitatively. With the increase of chemical potential, the quark-antiquark pair will melt. However, the chemical potential required for melting of the $\Sigma_u^-$ excited state is much lower.
In Fig.~{\ref{fig5}}, we calculate the regularized pseudopotentials of both the ground state and excited state for different values of chemical potential $\mu$. At small $L$, the $\Sigma_g^+$ pseudopotentials for any chemical potential are in the Coulomb-like form, while the $\Sigma_u^-$ hybrid pseudopotentials are in a quadratic function form of $L$. At long ranges, they both exhibit linear behavior. Reviewing Fig.~{\ref{fig3}}, we discover that the $\Sigma_u^-$ hybrid pseudopotentials at any temperature and chemical potential are in a quadratic function form at small $L$ and a linear form at large $L$. Compared with the impact of chemical potential on the $\Sigma_g^+$ pseudopotential, which is minimal at large distances and can be almost negligible at short distances, the chemical potential has a greater impact on the $\Sigma_u^-$ hybrid pseudopotential. An increase in chemical potential will reduce the $\Sigma_u^-$ hybrid pseudopotential.

\begin{figure}
\centering
\subfigure[.\, Separate distance of $\Sigma_g^+$ ground state]{
\label{fig4a}
\includegraphics[width=0.45\linewidth]{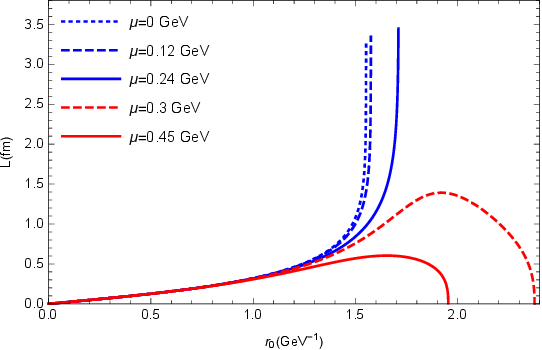}}
\subfigure[.\, Separate distance of $\Sigma_u^-$ excited state]{
\label{fig:subfig:4b}
\includegraphics[width=0.45\linewidth]{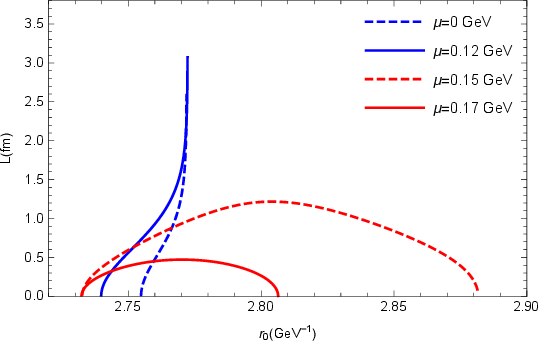}}
\caption{The separate distance $L$ as a function of $r_0$ corresponds to different values of chemical potential at the fixed temperature $T = 0.1\, {\rm GeV}$. }
\label{fig4}
\end{figure}

\begin{figure}
\centering
\subfigure[.$\Sigma_g^+$ pseudopotentials]{
\label{fig5a}
\includegraphics[width=0.45\linewidth]{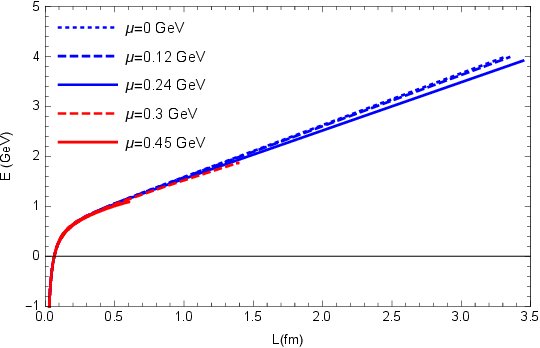}}
\subfigure[.\,$\Sigma_u^-$ hybrid pseudopotentials ]{
\label{fig:subfig:5b}
\includegraphics[width=0.45\linewidth]{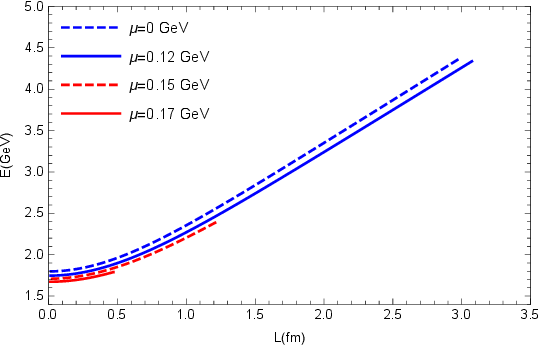}}
\caption{The regularized pseudopotentials of $\Sigma_g^+$ and $\Sigma_u^-$ as a function of separate distance $L$ at the fixed temperature $T = 0.1 \,{\rm GeV} $ for different values of chemical potential $\mu$. }
\label{fig5}
\end{figure}
To further investigate the influence of chemical potential on the hybrid pseudopotentials, we analyze the behavior of regularized pseudopotentials at a higher temperature, where quark-antiquark pair with any chemical potential could be melted, in Fig.~{\ref{fig6}}. Note that we have retained the knee points of the curves which exactly represent the critical point for melting, and its abscissa represents the screen distance of the quark-antiquark pair.
Fig.~{\ref{fig6a}} represents the pseudopotentials of $\Sigma_g^+$ at a higher temperature $T = 0.15\; {\rm GeV} $ and different values of chemical potential $\mu= 0.1\; {\rm GeV}$,\, $ 0.2\; {\rm GeV}$,\, $0.3\; {\rm GeV}$. When the chemical potential increases, the screen distance will decrease but the $\Sigma_g^+$ pseudopotentials hardly change. At such a higher temperature, the $\Sigma_g^+$ pseudopotentials at any value of $\mu$ are Coulomb-like before the quark-antiquark pair melts. Meanwhile, we plot the $\Sigma_u^-$ hybrid pseudopotentials curves at the temperature $T = 0.11 \,{\rm GeV}$ in Fig.~{\ref{fig:subfig:6b}}. It can be found that the $\Sigma_u^-$ hybrid pseudopotentials for any chemical potential seem to display as quadratic functions of separate distance before they melt. The screen distance and the hybrid pseudopotentials of $\Sigma_u^-$ significantly decrease as the chemical potential increases.
\begin{figure}
\centering
\subfigure[.$\Sigma_g^+$ pseudopotentials at $T = 0.15\, {\rm GeV}$]{
\label{fig6a}
\includegraphics[width=0.45\linewidth]{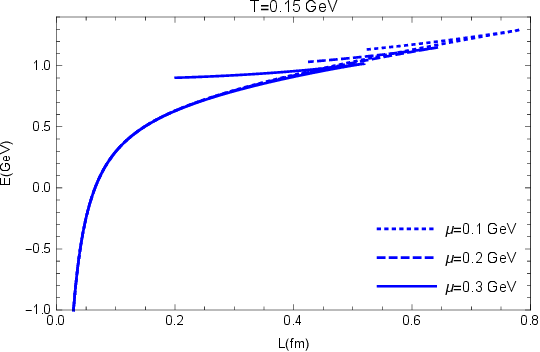}}
\subfigure[.\,$\Sigma_u^-$ hybrid pseudopotentials at $T = 0.11\, {\rm GeV}$]{
\label{fig:subfig:6b}
\includegraphics[width=0.45\linewidth]{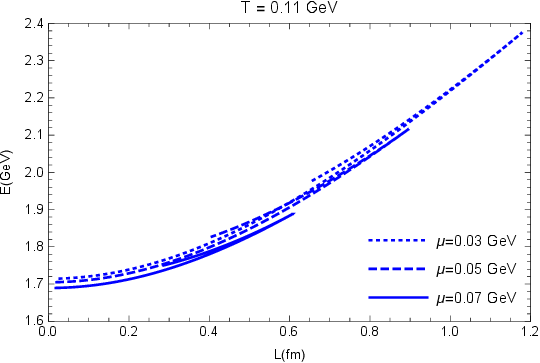}}
\caption{The regularized pseudopotentials corresponding to $\Sigma_g^+$ and $\Sigma_u^-$ at higher temperature . Here we have retained the knee points of the curves that represent the critical points for melting. The lower parts of these curves represent the pseudopotentials. (a) $\Sigma_g^+$ pseudopotentials at temperature $T = 0.15\, {\rm GeV} $ and different chemical potential $\mu= 0.1 \,{\rm GeV}$, \, $ 0.2 \,{\rm GeV}$, \, $0.3 \,{\rm GeV}$; (b) $\Sigma_u^-$ hybrid pseudopotentials at $T = 0.11\, {\rm GeV} $ and $\mu= 0.03 \,{\rm GeV}$, \, $ 0.05\, {\rm GeV}$, \, $0.07 \,{\rm GeV}$.}
\label{fig6}
\end{figure}

At last, the melting diagram of the quark-antiquark pair in both $\Sigma_g^+$ ground state and $\Sigma_u^-$ excited state are drawn in Fig.~{\ref{fig7a}}. The melting line of $\Sigma_u^-$ is lower than $\Sigma_g^+$. It can be inferred from the melting diagram that $\Sigma_g^+$ ground state is more stable than $\Sigma_u^-$ excited state.

\begin{figure}
\centering
{\includegraphics[width=0.50\linewidth]{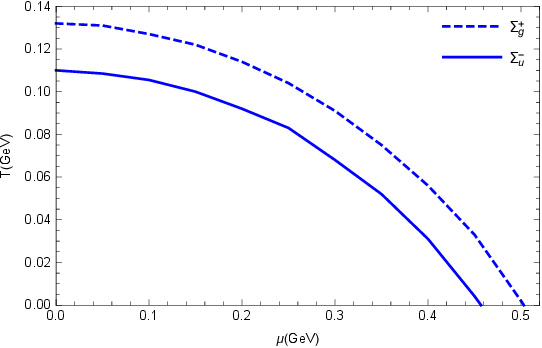}}
\caption{The melting diagram of quark-antiquark pair corresponding to $\Sigma_g^+$ ground state and $\Sigma_u^-$ excited state in the $ T - \mu$ plane. }
\label{fig7a}
\end{figure}

\section{Conclusion and discussion}
Heavy ion collisions can generate extremely high temperatures and densities in a very short amount of time. We mainly discussed the exotic hybrid pseudopotentials at finite temperature and chemical potential. To do so, we use a string configuration for the $\Sigma_u^-$ hybrid meson has been proposed by Oleg Andreev in Ref.~\cite{Andreev:2012mc, Andreev:2012hw}. In this paper, we focus on the influence of temperature and chemical potential on $\Sigma_g^+$ pseudopotentials and $\Sigma_u^-$ hybrid pseudopotentials. At short distances, the $\Sigma_u^-$ hybrid pseudopotential at finite temperature and chemical potential is no longer a Coulomb-like behavior but seems to exhibit a quadratic function of separate distance. It behaves as linear at long ranges, just like $\Sigma_g^+$ pseudopotential. It is a significant difference between the excited state and the ground state. In addition, we find that the screen distance and hybrid pseudopotential of $\Sigma_u^-$ significantly decrease as the temperature and chemical potential increase. However, temperature and chemical potential effects have a minor impact on $\Sigma_g^+$ pseudopotentials. At last, the melting diagram in the $ T - \mu$ plane implies that the quark-antiquark pair in the excited state is more likely to melt.

We provide a way to understand the $\Sigma_u^-$ hybrid pseudopotentials at finite temperature and chemical potential in this paper. Therefore the hybrid (pseudo)potentials of other hybrid mesons and baryons deserve further study. Moreover, we will consider the impact of some extreme conditions(e.g., rotating matter and magnetic field, etc.) on the hybrid (pseudo)potentials.

\section*{ACKNOWLEDGMENTS}

This work is supported by the National Natural Science Foundation of China (NSFC) under Grants No. 12405154, 12005056.

\section*{Data Availability Statement}
This manuscript has no associated data or the data will not be deposited.

\end{widetext}
%

\bibliographystyle{plain}
\bibliographystyle{h-physrev5}
\bibliography{zladscft_refs}

\begin{thebibliography}{10}

\bibitem{Fischler:1977yf}
W.~Fischler,
\newblock Nucl. Phys. B {\bf 129}, 157 (1977).

\bibitem{Brown:1979ya}
L.~S. Brown and W.~I. Weisberger,
\newblock Phys. Rev. D {\bf 20}, 3239 (1979).

\bibitem{Bali:1997am}
G.~S. Bali, K.~Schilling, and A.~Wachter,
\newblock Phys. Rev. D {\bf 56}, 2566 (1997), arXiv:hep-lat/9703019.

\bibitem{Baker:1997bg}
M.~Baker, J.~S. Ball, and F.~Zachariasen,
\newblock Phys. Rev. D {\bf 56}, 4400 (1997), arXiv:hep-ph/9705207.

\bibitem{Bali:2000gf}
G.~S. Bali,
\newblock Phys. Rept. {\bf 343}, 1 (2001), arXiv:hep-ph/0001312.

\bibitem{Takahashi:2002bw}
T.~T. Takahashi, H.~Suganuma, Y.~Nemoto, and H.~Matsufuru,
\newblock Phys. Rev. D {\bf 65}, 114509 (2002), arXiv:hep-lat/0204011.

\bibitem{Ratti:2005jh}
C.~Ratti, M.~A. Thaler, and W.~Weise,
\newblock Phys. Rev. D {\bf 73}, 014019 (2006), arXiv:hep-ph/0506234.

\bibitem{Bicudo:2007xp}
P.~Bicudo, M.~Cardoso, and O.~Oliveira,
\newblock Phys. Rev. D {\bf 77}, 091504 (2008), arXiv:0704.2156.

\bibitem{Luscher:2010iy}
M.~L\"uscher,
\newblock JHEP {\bf 08}, 071 (2010), arXiv:1006.4518,
\newblock [Erratum: JHEP 03, 092 (2014)].

\bibitem{Juge:2002br}
K.~J. Juge, J.~Kuti, and C.~Morningstar,
\newblock Phys. Rev. Lett. {\bf 90}, 161601 (2003), arXiv:hep-lat/0207004.

\bibitem{Koma:2017hcm}
Y.~Koma and M.~Koma,
\newblock Phys. Rev. D {\bf 95}, 094513 (2017), arXiv:1703.06247.

\bibitem{Muller:2019joq}
L.~M\"uller, O.~Philipsen, C.~Reisinger, and M.~Wagner,
\newblock Phys. Rev. D {\bf 100}, 054503 (2019), arXiv:1907.01482.

\bibitem{Brambilla:2021wqs}
N.~Brambilla {\em et~al.},
\newblock Phys. Rev. D {\bf 105}, 054514 (2022), arXiv:2106.01794.

\bibitem{Schlosser:2021hed}
C.~Schlosser and M.~Wagner,
\newblock SciPost Phys. Proc. {\bf 6}, 009 (2022), arXiv:2112.01911.

\bibitem{Schlosser:2021wnr}
C.~Schlosser and M.~Wagner,
\newblock Phys. Rev. D {\bf 105}, 054503 (2022), arXiv:2111.00741.

\bibitem{Schlosser:2022twt}
C.~Schlosser, S.~K\"ohler, and M.~Wagner,
\newblock PoS {\bf LATTICE2022}, 083 (2023), arXiv:2211.00489.

\bibitem{Philipsen:2013ysa}
O.~Philipsen and M.~Wagner,
\newblock Phys. Rev. D {\bf 89}, 014509 (2014), arXiv:1305.5957.

\bibitem{Bala:2020tdt}
D.~Bala and S.~Datta,
\newblock Phys. Rev. D {\bf 103}, 014512 (2021), arXiv:2009.00773.

\bibitem{Maldacena:1997re}
J.~M. Maldacena,
\newblock Adv. Theor. Math. Phys. {\bf 2}, 231 (1998), arXiv:hep-th/9711200.

\bibitem{Gubser:1998bc}
S.~S. Gubser, I.~R. Klebanov, and A.~M. Polyakov,
\newblock Phys. Lett. B {\bf 428}, 105 (1998), arXiv:hep-th/9802109.

\bibitem{Witten:1998qj}
E.~Witten,
\newblock Adv. Theor. Math. Phys. {\bf 2}, 253 (1998), arXiv:hep-th/9802150.

\bibitem{Maldacena:1998im}
J.~M. Maldacena,
\newblock Phys. Rev. Lett. {\bf 80}, 4859 (1998), arXiv:hep-th/9803002.

\bibitem{Aharony:1999ti}
O.~Aharony, S.~S. Gubser, J.~M. Maldacena, H.~Ooguri, and Y.~Oz,
\newblock Phys. Rept. {\bf 323}, 183 (2000), arXiv:hep-th/9905111.

\bibitem{Rey:1998bq}
S.-J. Rey, S.~Theisen, and J.-T. Yee,
\newblock Nucl. Phys. B {\bf 527}, 171 (1998), arXiv:hep-th/9803135.

\bibitem{Brandhuber:1998bs}
A.~Brandhuber, N.~Itzhaki, J.~Sonnenschein, and S.~Yankielowicz,
\newblock Phys. Lett. B {\bf 434}, 36 (1998), arXiv:hep-th/9803137.

\bibitem{Andreev:2006eh}
O.~Andreev and V.~I. Zakharov,
\newblock Phys. Lett. B {\bf 645}, 437 (2007), arXiv:hep-ph/0607026.

\bibitem{Fadafan:2012qy}
K.~B. Fadafan and E.~Azimfard,
\newblock Nucl. Phys. B {\bf 863}, 347 (2012), arXiv:1203.3942.

\bibitem{Lee:2014gna}
B.-H. Lee, C.~Park, and S.~Nam,
\newblock JHEP {\bf 05}, 011 (2015), arXiv:1412.3097.

\bibitem{Chen:2017lsf}
X.~Chen, S.-Q. Feng, Y.-F. Shi, and Y.~Zhong,
\newblock Phys. Rev. D {\bf 97}, 066015 (2018), arXiv:1710.00465.

\bibitem{Jiang:2022zbt}
J.-J. Jiang, Y.-Z. Xiao, J.~Qin, X.~Li, and X.~Chen,
\newblock Chin. Phys. C {\bf 47}, 013106 (2023), arXiv:2212.03541.

\bibitem{Chakraborty:2012dt}
S.~Chakraborty and N.~Haque,
\newblock Nucl. Phys. B {\bf 874}, 821 (2013), arXiv:1212.2769.

\bibitem{Finazzo:2014rca}
S.~I. Finazzo and J.~Noronha,
\newblock JHEP {\bf 01}, 051 (2015), arXiv:1406.2683.

\bibitem{Gursoy:2020kjd}
U.~G\"ursoy, M.~J\"arvinen, G.~Nijs, and J.~F. Pedraza,
\newblock JHEP {\bf 03}, 180 (2021), arXiv:2011.09474.

\bibitem{Zhou:2020ssi}
J.~Zhou, X.~Chen, Y.-Q. Zhao, and J.~Ping,
\newblock Phys. Rev. D {\bf 102}, 086020 (2020), arXiv:2006.09062.

\bibitem{Zhou:2021sdy}
J.~Zhou, X.~Chen, Y.-Q. Zhao, and J.~Ping,
\newblock Phys. Rev. D {\bf 102}, 126029 (2021).

\bibitem{Zhou:2023qtr}
J.~Zhou, S.~Zhang, J.~Chen, L.~Zhang, and X.~Chen,
\newblock Phys. Lett. B {\bf 844}, 138116 (2023), arXiv:2310.15609.

\bibitem{Andreev:2008tv}
O.~Andreev,
\newblock Phys. Rev. D {\bf 78}, 065007 (2008), arXiv:0804.4756.

\bibitem{Andreev:2015riv}
O.~Andreev,
\newblock Phys. Rev. D {\bf 93}, 105014 (2016), arXiv:1511.03484.

\bibitem{Andreev:2015iaa}
O.~Andreev,
\newblock Phys. Lett. B {\bf 756}, 6 (2016), arXiv:1505.01067.

\bibitem{Andreev:2019cbc}
O.~Andreev,
\newblock Phys. Lett. B {\bf 804}, 135406 (2020), arXiv:1909.12771.

\bibitem{Andreev:2020xor}
O.~Andreev,
\newblock JHEP {\bf 05}, 173 (2021), arXiv:2007.15466.

\bibitem{Andreev:2021bfg}
O.~Andreev,
\newblock Phys. Rev. D {\bf 104}, 026005 (2021), arXiv:2101.03858.

\bibitem{Andreev:2021eyj}
O.~Andreev,
\newblock Phys. Rev. D {\bf 105}, 086025 (2022), arXiv:2111.14418.

\bibitem{Andreev:2022cax}
O.~Andreev,
\newblock Phys. Rev. D {\bf 106}, 066002 (2022), arXiv:2205.12119.

\bibitem{Chen:2021bkc}
X.~Chen, B.~Yu, P.-C. Chu, and X.-h. Li,
\newblock Chin. Phys. C {\bf 46}, 073102 (2022), arXiv:2112.06234.

\bibitem{Yu:2023hzl}
B.~Yu, X.~Guo, X.~Chen, and X.-H. Li,
\newblock Phys. Rev. D {\bf 108}, 066007 (2023), arXiv:2305.19091.

\bibitem{Jiang:2023lmj}
J.-J. Jiang, X.~Chen, J.~Qin, and M.~A. Martin~Contreras,
\newblock (2023), arXiv:2310.04983.

\bibitem{Liu:2023hoq}
X.~Liu, J.-J. Jiang, X.~Chen, M.~Fujita, and A.~Watanabe,
\newblock (2023), arXiv:2310.08146.

\bibitem{Isgur:1984bm}
N.~Isgur and J.~E. Paton,
\newblock Phys. Rev. D {\bf 31}, 2910 (1985).

\bibitem{McNeile:2002en}
C.~McNeile,
\newblock Nucl. Phys. A {\bf 711}, 303 (2002), arXiv:hep-lat/0207001.

\bibitem{Juge:1997nc}
K.~J. Juge, J.~Kuti, and C.~J. Morningstar,
\newblock Nucl. Phys. B Proc. Suppl. {\bf 63}, 326 (1998),
  arXiv:hep-lat/9709131.

\bibitem{Michael:1998tr}
C.~Michael,
\newblock Nucl. Phys. A {\bf 655}, 12 (1999), arXiv:hep-ph/9810415.

\bibitem{Michael:2003xg}
C.~Michael,
\newblock (2003), arXiv:hep-ph/0308293.

\bibitem{Bicudo:2008na}
P.~Bicudo, M.~Cardoso, and O.~Oliveira,
\newblock PoS {\bf LATTICE2008}, 151 (2008), arXiv:0811.0535.

\bibitem{Wolf:2014tta}
P.~Wolf and M.~Wagner,
\newblock J. Phys. Conf. Ser. {\bf 599}, 012005 (2015), arXiv:1410.7578.

\bibitem{Meyer:2015eta}
C.~A. Meyer and E.~S. Swanson,
\newblock Prog. Part. Nucl. Phys. {\bf 82}, 21 (2015), arXiv:1502.07276.

\bibitem{Reisinger:2017btr}
C.~Reisinger, S.~Capitani, O.~Philipsen, and M.~Wagner,
\newblock EPJ Web Conf. {\bf 175}, 05012 (2018), arXiv:1708.05562.

\bibitem{Capitani:2018rox}
S.~Capitani, O.~Philipsen, C.~Reisinger, C.~Riehl, and M.~Wagner,
\newblock Phys. Rev. D {\bf 99}, 034502 (2019), arXiv:1811.11046.

\bibitem{Borgs:1985qh}
C.~Borgs,
\newblock Nucl. Phys. B {\bf 261}, 455 (1985).

\bibitem{Manousakis:1986jh}
E.~Manousakis and J.~Polonyi,
\newblock Phys. Rev. Lett. {\bf 58}, 847 (1987).

\bibitem{Andreev:2012hw}
O.~Andreev,
\newblock Phys. Rev. D {\bf 87}, 065006 (2013), arXiv:1211.0930.

\bibitem{Andreev:2006ct}
O.~Andreev and V.~I. Zakharov,
\newblock Phys. Rev. D {\bf 74}, 025023 (2006), arXiv:hep-ph/0604204.

\bibitem{Andreev:2024orz}
O.~Andreev,
\newblock Phys. Rev. D {\bf 109}, 106001 (2024), arXiv:2402.09026.

\bibitem{Andreev:2012mc}
O.~Andreev,
\newblock Phys. Rev. D {\bf 86}, 065013 (2012), arXiv:1207.1892.

\bibitem{Co:2021lkc}
R.~T. Co {\em et~al.},
\newblock JHEP {\bf 09}, 116 (2022), arXiv:2108.09299.

\bibitem{Witten:1998xy}
E.~Witten,
\newblock JHEP {\bf 07}, 006 (1998), arXiv:hep-th/9805112.

\bibitem{Andreev:2007zv}
O.~Andreev,
\newblock Phys. Rev. D {\bf 76}, 087702 (2007), arXiv:0706.3120.

\bibitem{Andreev:2007rx}
O.~Andreev,
\newblock Phys. Lett. B {\bf 659}, 416 (2008), arXiv:0709.4395.

\end{thebibliography}
\end{document}